\documentstyle{mn}
\newcommand{\be}{\begin{equation}}
\newcommand{\ee}{\end{equation}}
\title{The Pulsar Kick Velocity Distribution}
\author[B. M. S. Hansen \& E. S. Phinney]
{Brad M. S. Hansen$^{1,2 }$\thanks{email:{\bf hansen@cita.utoronto.ca}} \& E.
Sterl Phinney$^2$ \\
$^1$ Canadian Institute for Theoretical Astrophysics, University of Toronto, \\
\hspace{2mm} 60 St. George Street, Toronto, ON  M5S 3H8, Canada \\
$^2$ Theoretical Astrophysics, 130-33, California Institute of Technology,
Pasadena, CA,  91125, USA}
\date{5 August 1997}

\pagerange{\pageref{firstpage}--\pageref{lastpage}}
\pubyear{1997}
\begin{document}

\maketitle

\label{firstpage}

\begin{abstract}
We analyse the sample of pulsar proper motions, taking detailed account of
the selection effects of the original surveys. We
treat censored data
using survival statistics. From a comparison of our results with Monte Carlo
simulations,
 we find that the mean birth speed of a
pulsar is $\sim 250-300 \rm \, km \, s^{-1}$, rather than the 450~km~s$^{-1}$
found
by Lyne \& Lorimer (1994). The resultant distribution is consistent with a
maxwellian
with dispersion $\rm \sigma_v = 190 \rm \, km \, s^{-1}$.
 Despite the large birth velocities, we find that
the pulsars with long characteristic ages show the asymmetric drift, indicating
that
they are dynamically old. These pulsars may result from the low velocity tail
of the younger population, although modified by their origin in binaries and by
evolution in the galactic potential.
\end{abstract}


\section{Introduction}
The fact that pulsars have velocities much in excess of those of ordinary
stars (a subset of whom are presumably the pulsar progenitors) has been known
almost since their discovery (Minkowski 1970; Trimble 1971; Lyne, Anderson \&
Salter 1982).
 The origin of these velocities is not so clear. One possibility is
that they result from the disruption of a binary population (Gott,
Gunn \& Ostriker 1970;
Iben \& Tutukov 1996), leaving the pulsar
with a velocity characteristic of the orbital velocity of the progenitor
 in the binary. The problem
with this scenario is that it has trouble explaining the largest observed
velocities
(e.g. Phinney and Kulkarni 1994; see, however, Iben \& Tutukov 1996).
 Another possibility is that the pulsar acquired its velocity
from an asymmetric supernova collapse. This hypothesis has
recently been bolstered by the observations of binary pulsar PSR J0045-7319, in
which
the observed spin-orbit precession and orbital period decay are strong
arguments for
a natal kick (Lai, Bildsten \& Kaspi 1995; Lai 1996).

 Lyne \& Lorimer (1994) have analysed the known sample of pulsar velocities
in the light of recent proper motion studies (Harrison,Lyne \& Anderson 1993)
as
well as the new pulsar distance scale due to Taylor and Cordes (1993). They
conclude
that pulsars are born with a mean speed of $\sim 450 \rm \, km \, s^{-1}$.
Although Lyne and
Lorimer restrict their sample to those younger than $4 \times 10^6$ years to
avoid
the `vertical leakage' selection effect pointed out by
 Helfand \& Tademaru (1977) and Cordes (1986), they did not treat the selection
effects that result from the flux limits of the pulsar surveys or the limiting
accuracy of proper motion determinations. We shall attempt to do that here.
Recently,
 Iben and Tutukov (1996) have also addressed this question, but in a less
systematic fashion
than what we propose to use in this paper.

We restrict our analysis to only those pulsars with velocities determined
by proper motion measurements. While scintillation data have been used to
estimate velocities
 (Cordes 1986; Harrison and Lyne 1993) to within a factor of 2, we prefer to
keep our
sample as homogeneous as possible, and so we exclude these data.
We use the properties of well-known pulsar surveys to estimate the $V/V_{\rm
max}$ correction
(Schmidt 1968)
for each pulsar, making use of survival statistics (Feigelson \& Nelson 1985)
to treat those data with upper bounds (section~\ref{survstat}). This allows us
to estimate the
two dimensional velocity distribution of the observed pulsar population and
thus the kick velocity
distribution taking into account the differential galactic rotation in section
\ref{Kicks}.

\section{The Proper Motion Distribution}
\label{veldis}
\subsection{Selection Effects}
The Princeton pulsar database (Taylor, Manchester \& Lyne 1993) now
 contains $\sim$ 800 pulsars, 101 of
whom have measured proper motions or upper limits. The number of surveys
responsible for
this profusion is also gradually increasing in size (in excess of 15). To do a
proper
treatment of the selection effects for the full proper motion sample would then
require
modelling the selection effects of a significant number of these surveys.
Luckily, we note that most of the
proper motion pulsars were detected in the earlier surveys. By restricting
ourselves
to those pulsars detected in the Molonglo 2  (Manchester et al. 1978) and Green
Bank/NRAO 1,2 and 3 surveys
(Damashek et al. 1978, Dewey et al. 1985, Stokes et al. 1985, Stokes et al.
1986),
we are left with 86 out of 101 pulsars. We note that 12 of the 15 pulsars left
out have
P$<0.1$ s (and only one of our restricted sample satisfies this criterion),
 which is not surprising, since  many of the later surveys focussed on finding
faster spinning
pulsars. In performing this cut, we lose one young pulsar and all but one of
the pulsars
with characteristic ages greater than 1 Gyr. This also means we are not
affected by the
possibly different evolutionary histories of millisecond pulsars (in particular
the
influence of binaries). We shall omit the one
old binary pulsar (PSR 0655+64) which does fall into our sample as well.
Figure~\ref{select}
shows the distribution of the included and excluded pulsars as a function of
period and
velocity.


The observed pulsar sample suffers from two obvious selection effects, due to
flux limits
and proper motion limits respectively. Figure~\ref{Lv} shows the distribution
of inferred
luminosities and transverse velocities. The lack of faint, fast pulsars (upper
left
corner)  and bright, slow
pulsars (lower right corner) is evident.


To correct for this bias, we need to weight the pulsars according to the
maximum volume
in which they could have been detected, i.e. using a $V_{\rm max}$ weighting.
To do this we need to consider the detection efficiency of the various
pulsar surveys. After  Narayan (1987) (see also Dewey et al. 1984), the minimum
flux detectable is
\be S_{\rm min} = S_{\rm 0} \frac{ \left(T_{\rm rec} + T_{\rm
sky}\right)}{T_{\rm 0}} \sqrt{
\frac{P W}{W_{\rm e}\left(P-W\right)}}, \ee
where $ T_{\rm rec}$ and $ T_{\rm sky}$ are
the system and sky noise temperatures, $ S_{\rm 0}$ is the flux normalization,
 and W is the measured
pulse width. $ W_{\rm e}$ is the intrinsic pulse width, which is broadened
because of
sampling, dispersion and scattering so that
\be W^2 = W_{\rm e}^2 + \tau_{\rm samp}^2 + \tau_{DM}^2 + \tau_{\rm scatt}^2.
\ee
Thus we calculate the minimum flux for a given survey in a particular
direction.
 The parameters
describing each survey were taken from  Narayan (1987)\footnote{Narayan's
formula  for the sky
background temperature in equation (3.5)  contains a typographical error. The
factor $ \left[1+(b/3)^2\right]$ should be
in the denominator.} and  Stokes et al. (1986).
We used the updated
electron distribution model of Taylor and Cordes (1993) to calculate the
dispersion and scatter
broadening along a given line of sight. This accounts for the bias due to the
flux limits.
No such simple model exists for treating the  proper motion limits. This is
because the
accuracy of a given proper motion measurement depends on the vagaries of the
distribution
of background radio sources near the pulsar position on the sky (Harrison, Lyne
\&
Anderson 1993). As a crude model of this, we fit the distribution of proper
motion errors
in Harrison et al. using the distribution $\rm p(\mu) = exp(-\mu/10.5\, mas \,
yr^{-1})$.
In the $ V_{\rm max}$
calculations to follow, the proper motion cutoff is randomly selected from this
distribution
for each line of sight and .
 We also used simple limits of 5 and 2 $\rm mas \, yr^{-1}$ as a test of the
importance
of this selection effect. This introduces
a variation of $\sim \rm 10 \, km \, s^{-1}$ in the mean velocity of the young
sample and about
$30 \rm \, km \, s^{-1}$ in the old sample, indicating that most of the $
V_{max}$ are
flux-limited rather than proper motion limited. Iben and Tutukov (1996) have
also taken account of
this selection effect.

For each pulsar, we randomly place it in different directions and
at
different distances with respect to the observer and calculate whether or not
it would
be detectable with it's known luminosity and transverse velocity (uncorrected
for the local standard
of rest)
in any of the surveys we consider. Thus, using this Monte Carlo integration
procedure, we determine the volume within which each pulsar could have been
detected.
These $V_{\rm max}$ values determine the relative weights of each of the
pulsars in the
corrected sample.

A possible source of concern with this procedure is illustrated by
Figure~\ref{Slim}.
The analytic flux limits do not describe the complications of the true
detection
limits perfectly (interstellar scintillation makes the apparent flux vary).
 In Figure~\ref{Slim} we find four pulsars detected by the Molonglo
survey that
lie below the analytically described detection threshold for that survey. This
will reduce the
weight accorded to these pulsars. To estimate the impact of this error on our
results,
we repeated the analysis with these pulsars artificially `brightened' to meet
the
flux limit expression. The mean proper motion we infer for the young pulsar
sample
increases by only 7 $\rm km \, s^{-1}$, so this is not a significant source of
error
for our analysis here.


The $V_{\rm max}$ correction is not without biases of its own. In particular,
weighting pulsars by their
 $V_{\rm max}$ presupposes that the real population is distributed uniformly
throughout the galactic
volume. However, the pulsars are born from a disk population with  a scale
height of about
150-450 pc (Narayan and Ostriker 1990). Thus, a population born with small
velocities will not expand
to fill as much of the spherical volume as a fast population. The above
analysis then overcorrects for the slow pulsars (see Helfand and Tademaru 1977;
Cordes 1986;
Lyne \& Lorimer 1994). In order to adjust for this, we consider the maximum
detectable volume to be limited in the vertical extent by the scale $Z_{\rm
max}$=$ V_t t_p$,
where $ t_p$ is the pulsar timing age. If this is
larger than $D_{\rm eq}=V_{\rm max}^{1/3}$, then there is no change in the
weight assigned to that pulsar, but, if $ D_{\rm eq} >
Z_{\rm max}$, then we assume that we see the edge of the distribution of
pulsars of this velocity, and reduce
the weight given to that pulsar accordingly(see Figure~\ref{Volume}).
Furthermore, for young pulsars, we set a lower limit on $ Z_{\rm max}$ of
450 pc, representative of the initial scale height.


In Figure~\ref{tvw} we show the distribution of velocities with age and the
relative weighting
of each pulsar.


\subsection{Survival Statistics}
\label{survstat}
The $ V_{\rm max}$ correction takes care of the selection effects, but we still
need to
account properly for those data which only have upper limits (`censored' data
in the
statistical lexicon). Of our 85 pulsars, 20
fall into this category. Using only those data with actual detections will bias
our
distribution to higher values as we will see below.

Following Feigelson and Nelson (1985), we use survival statistics to treat the
effects
of our censored data. In particular, we use the Kaplan-Meier estimator (Kaplan
and Meier
1958) to calculate the cumulative probability distribution of the transverse
velocities. However, we need to modify this method slightly to take account of
our $V_{\rm max}$
correction.

Consider our data to be the set  of n distinct values $ \{ x_{i} \}_{i=1}^{n}
$, where the $ x_i$
are ordered in the manner
$  x_1 > x_2 > \cdots > x_n $, where $ x_{i}$  can be either a detected value
or an upper bound.
Now, for a given value $t$, let
$$
 P_i = P\left[ t \leq x_{i+1} \left| t \leq x_i \right. \right],
$$
i.e. the conditional probability that $t$ is  less than $ x_{i+1}$ given that
it is less than $ x_i$.

Using this, we can calculate the probability that t is less than any given $
x_i$ ($i>1$) by
$$
 P\left[ t \leq x_{i} \right] = \Pi_{j=1}^{i-1} P\left[t \leq x_{j+1} \left| t
\leq x_j \right. \right].
$$
To calculate this, we need to estimate the $P_j$. To start, $ P_1 = 1$, since
all values are at most
as big as $ x_1$. For any other $ x_j$, if it is a detection, then there are $
n-j+1$ values at
most as large as $ x_{j}$, and all except $ x_{j}$ are also at least as large
as $
 x_{j+1}$. Thus, we
estimate
$$
 P_j = {n - j\over n-j+1} = 1 - { 1\over n-j+1}.
$$
If $ x_j$ is an upper bound, then $ P_j =1$ by following similar reasoning as
above. Thus, we have
that
$$
 P\left[t \leq x_{i} \right] = \Pi_{j=1}^{i-1} \left( 1 - {1 \over n-j+1}
\right)^{\delta_j },
$$
where
$ \delta_j=1$ if $ x_j$ is a detection, and $ \delta(j)=0$ if $ x_{j}$ is an
upper limit.
This is the Kaplan-Meier (1958) estimator of the distribution function.
In the case where we have ties in our data (i.e. more than one measurement at
the same value),
this becomes
$$
 P\left[t\right] = \Pi_{j,x_j>t} \left( 1-{d_j \over n_j} \right)^{\delta_j},
$$
where $ n_j =$ number of measurements $ \leq x_j$, and $ d_j$ is the number of
measurements at the
value $ x_j$.

We also want to calculate the mean of velocities. Since $p(x) dx = \Delta P$,
the mean of a
quantity $x$ is given by
$$
 <x> = \int_0^{1} x \, dP = \int_0^{\infty} P \, dx,
$$
where we have integrated by parts. Thus we estimate the mean by
$$
 < V_t > = \sum_{i=1}^{N} P\left[ x_i\right] \left( x_i - x_{i-1} \right),
$$
and we take $ x_0=0$.

To include a $V/V_{\rm max}$ weighting, we adjust the number of `counts' at
each value according to
the weights $w(j)$, and thus we have
$$
 {d_j \over n_j} = \sum_{i\leq j} {w_j \over w_i}.
$$
In calculating the error on this new estimator, we note
that we have artificially increased the total number of observations fed into
the sum and thus have
reduced the error by a factor $ \sqrt{N_{\rm W}}$, where $ N_{\rm W}$ is the
number of pulsars including the
weightings. We remove this bias by multiplying the errors by $ \sqrt{N_{\rm
W}/N}$ where $N$ is the
true number of points in the sample.

\subsection{The Corrected Distribution}
\label{corrected}


Figure~\ref{pdis} shows the cumulative probability distribution of observed
transverse velocities taking into account different levels of adjustment.
With no adjustments, the mean transverse velocity of the entire sample is
355~km~s$^{-1}$, in agreement
with the analysis of Lyne \& Lorimer (1994). The mean of the entire sample
including
all corrections is 195~km~s$^{-1}$.
However, if we restrict
 the sample to
pulsars with characteristic ages $< 10^7$ years, we obtain a mean transverse
velocity of 237~$ \pm$~49
$\rm km \, s^{-1}$ (51 pulsars). The mean velocity of the complementary sample
with ages $> 10^7$ years
is 193~$\pm 50$~$\rm km \, s^{-1}$ (35 pulsars). The similarity between this
and the overall sample
is due to the great weight assigned to old, faint and slow moving pulsars (see
also the discussion of
the Iben \& Tutukov results in section~\ref{discussion}).
 In Figure~\ref{pdis} we see that the distribution for
the old pulsars does have a larger low velocity tail, as one might expect. If
we place the
velocity cutoff at $4 \times 10^6$ years, we find 226~$\pm$~71~$\rm km \,
s^{-1}$ for the 36 young
pulsars and 198~$\pm$~53~$\rm km \, s^{-1}$ for the older pulsars.

Our analysis of the proper motion errors neglects the effects of the
differential rotation or
corrections to the pulsar local standard of rest. However, these are not
important corrections for
two reasons. The first is that the pulsar sample is largely restricted to
distances $< 3 \rm kpc$, so
that these corrections are, at most, $\sim 2 B \times \rm 3 kpc \sim 75 km \,
s^{-1}$,
 where B is Oort's B constant. At this level it might still have observable
consequences, but the
second reason is that
 the $V_{\rm max}$ values are determined
largely by the flux limits, rather than the proper motion limits. Thus, these
effects are only likely
to become important when observers attempt to extend the proper motion sample
to greater distances.

\subsection{The Asymmetric Drift}

A reduction in the mean velocity for old pulsars has been noted
before by several authors (Lyne \& Lorimer 1994; Nice and Taylor 1995;
 Camilo, Nice \& Taylor
1993). We should note, however, that when we refer to ``old'' pulsars, we refer
to those
with characteristic ages less than $10^9$ years, i.e. we don't consider
millisecond pulsars because
of the increased complexity of treating their selection effects. Nevertheless,
in a completely
model-independent way, we can demonstrate that these pulsars are old in a
dynamical sense,
because they show
the effects of the asymmetric drift (e.g. Mihalas
and Binney 1981).  Nice and Taylor (1995) have pointed out that the millisecond
pulsar
population might possess
this property, but it appears to be true for all pulsars with spin down ages $>
10^7$ years.
 This is shown in Figure~\ref{asd}. The effect has its origin in the fact that
any virialised population with a significant radial velocity dispersion will
rotate about the galactic
centre more slowly than the local circular speed (Mihalas and Binney 1981).

 To calculate this we restrict ourselves
only to those pulsars with well-determined proper motions (since large enough
error bars can
reverse the sign of the transverse velocity). However, because we will not
treat selection
effects in this case, we shall use all the proper motion pulsars that satisfy
this and
subsequent criteria, including millisecond pulsars.
 We consider a cartesian coordinate system with
origin at the sun, with  positive x pointing radially outward and positive y
pointing in the
direction of $\ell = 270^{\circ}$. We consider the y-components of the
transverse velocity,
which measures the approximate azimuthal component of the pulsar velocity with
respect to the
sun (strictly speaking, this should be done for each pulsar in its individual
local standard
of rest, but this rough approach demonstrates our result sufficiently well and
remains the same
if we reduce the sample radius). We also exclude pulsars with $d>$ 6 kpc and
within 20$^{\circ}$ of
$\ell$=90 or 270 (where transverse velocities are primarily radial in the
galactic frame)
 and of $b = \pm 90^{\circ}$. This leaves
us with 37 pulsars with characteristic ages from $10^3$ - $10^{10}$ years. The
signature of the asymmetric
drift is thus an excess of positive $ V_{\rm y}$.
Indeed, this is seen to
striking effect in figure~\ref{asd}, where less than 10\% of the pulsars with
characteristic
ages older than $4 \times 10^6$
years have negative $V_{\rm y}$. This indicates that these pulsars must be at
least $10^7$ years
old, which supports the use of the age cutoff in section~\ref{corrected}. It
also provides a
lower limit for the magnetic field/torque decay time, consistent with that
found by recent analyses (Bhattacharya
et al 1992, Hartman et al 1996).


\section{The Kick Distribution}
\label{Kicks}

In section~\ref{veldis}, we derived the corrected proper motion distribution
appropriate to
a volume limited sample. Since pulsars receive their kicks in a rest frame
rotating about the
galactic centre and we are interested in the low
velocity tail, we need to consider the effect of differential galactic
rotation.

Using different initial velocity distributions, we project the velocities along
random lines of
sight in a sphere of 3 kpc radius about the observer and compare the inferred
proper motion distribution
with what we observe. This is shown in Figure~\ref{vdis}.
We find the distribution is
consistent with a maxwellian distribution of kick velocities with velocity
dispersion
of $\sim$ 190 km \, s$^{-1}$ (corresponding to a 3-D mean of $\sim 300 \rm \,
km \, s^{-1}$). However, it is not consistent with the Lyne and Lorimer proper
motion
distribution or
 the earlier form suggested by Paczynski (1990)\footnote{p(u)=$4/\pi(1+u^2)^2$,
where u=V/V$_*$ and
V$_*$ is some normalisation constant. We have chosen V$_* = 500 {\rm
km.s^{-1}}$ to be consistent
with the observed mean of the distribution.}. Neither of these
analyses included an extensive discussion of selection effects. Thus, our best
estimate for the kick velocity
distribution is
\be p(V_{\rm k}) = \sqrt{\frac{2}{\pi}} \frac{v^2}{\sigma_v^3} e^{-V_{\rm
k}^2/2 \sigma_v^2}
\label{pkick} \ee
with $\rm \sigma_v = 190 \, km \, s^{-1}$. However, there is a significant
degree of uncertainty about the
form of the distribution, in particular at the high end. As a demonstration of
this, we propose a second
extreme form which consists of two delta functions, at 250~$\rm km.s^{-1}$ and
1000~$\rm km.s^{-1}$, weighted such that 20 \% of the pulsars acquire the
higher velocity. Thus,
\begin{eqnarray} p(V_{\rm k}) & = & 0.8 \, \delta \left(V_{\rm k}-250 \, {\rm
km.s^{-1}}\right) + \nonumber \\
&&  0.2 \, \delta
\left( V_{\rm k} - 10^3 \, {\rm km.s^{-1}}\right), \end{eqnarray}
where $\delta(V-V_0)$ is the Dirac delta function.
Both forms are shown in Figure~\ref{vdis}.


\section{Long Term Evolution}
\label{Evolve}

\begin{table*}
\centering
\begin{minipage}{140mm}
\caption{Parameters of our Galactic Model \label{Numbers}}
\begin{tabular}{lcccc}
 & $a$ & $b$ & $r_{\rm c}$ &
 $M$  \\
 & (kpc) & (kpc) & (kpc) & ($\rm M_{\odot}$) \\
\hline
 Bulge & 0 & 0.277 &  &  1.12 $\times 10^{10}$ \\
 Disk & 3.7 & 0.20 & & 8.07 $\times 10^{10}$ \\
 Halo & & & 6.0 & 5.0 $\times 10^{10}$ \\
 \hline
\end{tabular}
\end{minipage}
\end{table*}

Our analysis above is concerned solely with those pulsars with characteristic
ages $< 10^7$ years.
To do the same for the older pulsar population will require the incorporation
of the selection
effects of more pulsar surveys, as well as the effect of the death line and
recycling of
pulsars in binaries. Nevertheless, if we assume that all pulsars originate from
a population with the velocity distribution (\ref{pkick}), we may examine the
long-term
evolution
of this population. We have performed
Monte Carlo simulations of such a population using a galactic potential from
Paczynski (1990).
The potential contains two terms of the form
\be
 \Phi_{i}(R,z) = \frac{-G M_{i}}{\left[ R^2 + \left( a_i + (z^2 + b_i^2)^{1/2}
\right)^2 \right]^{1/2}}
\ee
where $i$=1,2 represent the disk and bulge respectively. A third component, the
halo, is
represented by
\be \Phi_3(r) = -\frac{G M_{\rm c}}{2 r_{\rm c}} \left[\ln \left(1 +
\frac{r^2}{r_{\rm c}^2} \right)
+\frac{2 {\rm arctan} \left( \frac{r}{r_{\rm c}}\right)} {r/r_{\rm c}} \right],
 \ee
and $r = \left( R^2+z^2\right)^{1/2}$.
The various parameter values are given in Table~\ref{Numbers}.
The birth positions of the pulsars are distributed exponentially in both
galactocentric radius
($R$) and
disk height ($z$), with scale lengths of 4.5 and 0.075 kpc respectively. The
integration is performed
using the Burlisch-Stoer integration algorithm from  Press et al. (1992).


To recreate the observations we `observe' the pulsars from a galactocentric
radius of 8 kpc,
using a volume of radius 3 kpc about the observer (this distance is chosen to
approximately
recreate the volume sampled by Harrison, Lyne \& Anderson 1993). To make more
efficient use
of our simulations we can use 8 simultaneous observers spread equidistantly
around the circle
R = 8 kpc (see Figure~\ref{Observe}).
 As long as the observing volumes don't overlap, the observations from all
volumes
can be added together. We calculate the transverse velocities after subtracting
the circular
velocity of the observer and derive the observed velocity distribution as a
function of
time.


Figure~\ref{vevol} shows the evolution of the transverse velocity distribution
with time.
We can compare it to our corrected observed distributions, both for $ t < 10^7$
years.
We see that the $10^5$ year distribution confirms our analysis of
section~\ref{Kicks}.  and
similarly for $ t > 10^8$ years (see below).

An additional constraint on the evolved distribution is the magnitude of the
asymmetric
drift. For the kick distribution given by equation~(\ref{pkick}) the magnitude
is small ($\rm V_{\ell} < 10 \,
km \, s^{-1}$) to begin with, reaching a maximum of $\sim$ 80 $\rm km \,
s^{-1}$ after $10^8$ years
and then reaching an asymptotic value of $\sim 60$ $\rm km \, s^{-1}$ after
$10^9$ years. The
evolution of this quantity is shown in Figure~\ref{asd}. Again, we see we have
the correct
order of magnitude. Figure~\ref{simasd} demonstrates the magnitude of the
effect
(although it will be modified by the approach to the death line and binarity of
the
older pulsars) in our simulations as we show $V_x$ (radial in observer frame)
and $V_y$(transverse)
at different ages. Apart from the conspicuous appearance of the asymmetric
drift at later times,
we note also an excess of pulsars moving radially outwards at ages$\sim 10^7$
years. This is because
more pulsars are born at smaller galactocentric radii and the timescale for the
escaping
pulsars to travel a few kpc at $\sim 300$ $\rm km \, s^{-1}$ is a few $\times
10^7$ years. The smaller
signal makes this a more difficult effect to observe in the real pulsar
population.


As we have noted before, a proper comparison of our results with the older
pulsars
requires an analysis of the millisecond pulsar distribution. A treatment of the
evolution of the entire pulsar sample is beyond the scope of this paper, but we
can
answer the question whether the oldest pulsars have a velocity distribution
that is
a direct result of the kick distribution given by equation~(\ref{pkick}).
 To do this we model the
effects of the surveys undertaken at Molonglo (D'Amico et al. 1988), Jodrell
Bank (Biggs
\& Lyne 1992), Arecibo (Wolszczan 1991; Nice, Taylor \& Fruchter 1993; Camilo
et al.
 1993; Foster, Wolszczan \& Camilo 1993; Thorsett et al. 1993) and Parkes
(Johnston
et al. 1993; Bailes et al. 1994; Lorimer et al. 1995), although the $V_{\rm
max}$ was
dominated by the
 surveys from the latter two
telescopes. By restricting ourselves to pulsars with characteristic ages $>
10^8$ years,
 we
hope to determine the velocity distribution of the oldest pulsars.
Note that $\sim 40 \%$ of this sample are still long period (P $> 10 \, \rm
ms$)
pulsars. However, if we further restrict our sample to only millisecond
pulsars, we
obtain a similar result.
 Figure~\ref{vevol}
shows the  $V_{\rm max}$ weighted velocity distribution for this sample.
 We note that our derived distribution is strongly peaked between
60-100~$\rm km \, s^{-1}$, with the biggest weights attributed to the pulsars
B0655+64,
B1952+29 and J2322+2057, although 6 other pulsars also lie within this range as
well.

A striking feature of Figure~\ref{vevol} is that the old pulsars do not have a
velocity
distribution that comes from an unadulterated evolution of the initial velocity
distribution.
However this is not surprising, since
 many of the older pulsars
are in binaries. Thus, the distribution of centre of
mass velocities will be different from the kick velocities given to the neutron
star component.
Calculations of this effect have been performed
by numerous authors for different kick distributions
 (e.g. Dewey and Cordes 1986; Brandt and
Podsiadlowski 1995; Kalogera 1996) and introduce further
uncertainties into the problem through the initial distribution in orbital
period, companion
masses and pre-supernova helium star masses. Thus we shall defer detailed
 discussions to future work, although we note that the effect is indeed to
reduce the high
velocity tail. Another effect is that the low velocity end is also expected to
be depleted
due to the gravitational scattering off giant molecular clouds
 (Spitzer \& Schwarzschild 1951, 1953) or spiral arms (Barbanis \& Woltjer
1967).
This is a concern
for calculations of the observability of old neutron stars accreting from the
ISM (
because the accretion rate and thus luminosity $ \propto v^{-3}$), and
Madau \& Blaes (1994)
 have recently calculated this effect,
yielding a peak in the range 60-90 $\rm km \, s^{-1}$ after $10^{9-10}$ years (
Although they used the Narayan \& Ostriker (1990) kick distribution).

\section{Discussion}
\label{discussion}

Our approach above is designed to provide a robust estimate of
the characteristic birth velocities of pulsars. We have tested the procedure
with simple Monte Carlo
simulations using known distribution functions. We find that we can reproduce
the mean velocities to within
$\sim 30-50$ km \, s$^{-1}$, although the exact shape of the distribution at
the high
velocity end ($> 300 \rm \, km \, s^{-1}$) is not well constrained.
To better estimate the shape of the distribution will require more detailed
modelling and
the use of more of the pulsar population information,
such as was done by Narayan and Ostriker (1991) or Bhattacharya et al. (1991).

Another recent analysis by Iben and Tutukov (1996) finds a large birthrate of
very slow ($< 10 \rm
\, km \, s^{-1}$) pulsars. Their treatment neglects flux limits (although some
inferences are made on
the basis of a nearby sample only, their full analysis uses pulsars at all
distances),
 and treated the proper motion limits using various ad hoc analytic cutoffs.
They
also used no upper age cutoffs, so that their large birthrate of slow pulsars
was due to
3 old ($\rm t > 10^7$ years) pulsars, which acquired significant weight because
of their small
luminosities. When restricting ourselves to ages $<10^7$ years, we do not find
any
such low velocity tail. Furthermore, we find that the flux cutoff is
responsible for the
detection limit of more pulsars than the proper motion limit.

Some constraints on the nature of pulsar kicks can be obtained by analysing the
properties
of binaries containing neutron stars.
Recently, in the light of the results of Lorimer and Lyne (1994),
 Brandt and Podsiadlowski
(1995) analysed the effect of the revised kick distribution on the
post-supernova orbital
parameters of neutron star binaries. Their analysis indicated that an
isotropically distributed
kick velocity of 450 km~s$^{-1}$ was inconsistent with the eccentricity-orbital
period
distribution of the observed binary population. However, a kick velocity of 200
km~s$^{-1}$ was
perfectly consistent. Similarly, Wijers et al. (1992) obtained an upper limit
of 400
km~s$^{-1}$ for a characteristic kick velocity from an analysis of the
eccentricities
of the known double neutron star binaries. While these results are in good
agreement with
our analysis, we should note that
 Brandt \& Podsiadlowski showed that
the Lyne \& Lorimer distribution, despite it's high characteristic velocity,
 was also consistent with the period-eccentricity relation,
by virtue of it's large low velocity tail. Thus, the surviving binary
parameters do not provide
a means for distinguishing between the two distributions.

The most direct test of the pulsar kick velocities is to find the supernova
remnant
associated with the birth of a given pulsar. If the pulsar was born at the
centre of
a circularly symmetric supernova remnant, we may infer a proper motion from
it's displacement
 and thus obtain a velocity.
Much work has been devoted to this (Caraveo 1993; Frail, Goss \& Whiteoak
1994),
 and the results
indicate velocities significantly larger than the mean velocity derived by
other methods.
 In fact, the mean
velocity of the Frail et al. sample is 990~km~s$^{-1}$, with a median of
480~km~s$^{-1}$.
Whether or not this discrepancy with respect to our results
is real depends on the veracity of the various assumptions
used to infer a proper motion from a supernova association (see Kaspi 1996).
 Some of the problematic
assumptions discussed by Frail et al. include the difficulty of defining the
shape of a
remnant and the possible displacement of the supernova blast centre with
respect to the geometric centre of the remnant
(perhaps caused by expansion into an inhomogeneous surrounding medium). It is
illuminating, although
not conclusive, to
note that the two pulsars associated with supernova remnants which have
measured proper motions are the
Crab and Vela pulsars, with transverse velocities of 150 and 120 km~s$^{-1}$
respectively (In particular,
Frail et al. note that the methods used on other pulsar-remnant associations
would imply a velocity of
800 km~s$^{-1}$ for the Vela pulsar!)

A revised velocity distribution can possibly affect the results of statistical
analyses of
the pulsar population as a whole. Indeed, it may even contribute to an
explanation for the
conflicting claims concerning magnetic field decay.
 Narayan and Ostriker (1990) used a maxwellian distribution for their paper on
the pulsar
population as a whole. They used two different populations of pulsars and their
kick distribution was a function of magnetic field.
Their two populations had velocities that varied from 80-250 $\rm km \, s^{-1}$
for their S population
and from 20-100 $\rm km\, s^{-1}$ for their F population.
The analysis of Bhattacharya et al.
(1991) also used a maxwellian, but with a somewhat lower dispersion of 110
km~s$^{-1}$.
Our value lies a little above these characteristic values, consistent with the
fact that
the dispersion measure distances are now thought to be larger than those used
in the
above analyses. More recent analyses, such as that of Hartman (1996), attempt
to constrain the
velocity distribution itself using this approach. Results from such studies may
differ significantly
from ours because of the sundry additional assumptions required. In particular,
the Crab and
Vela pulsars acquire significant weight in pulsar current studies (see Phinney
\& Blandford 1981)
because they require significant luminosity  evolution to avoid a large
population of high field
objects near the death line. This could produce a larger low velocity tail than
in our case.
Similarly, inclusion of some of the more optomistic pulsar-supernova remnant
associations in the population synthesis can
result in a high velocity tail. Relative to such analyses, our results
represent a minimal model,
required to describe the observed proper motions under the assumption that the
death line does
not remove a significant fraction of pulsars before a spin-down age of $10^7$
years.

An interesting feature of our result is the lack of a pronounced low velocity
tail.
This has implications for the retention fraction of neutron stars in globular
clusters, as
well as the existence of some obviously low velocity pulsars with ages $> 10^7$
years.
Globular cluster central escape velocities are  $\leq$ 50 km~s$^{-1}$.
If all kick velocities are $\sim$ 250 km~s$^{-1}$, then no pulsars born from
isolated stars
are retained! If the distribution is
maxwellian, then the fraction retained is about 0.2\%, which is still extremely
low!
 Yet, the retention fraction of neutron stars is claimed to be of the order
of 10\%  (Phinney 1993) or higher (Hut and Verbunt 1983).
 However, it is
possible that binaries, either primordial or dynamically formed, could be
responsible
for the pulsars found in globular clusters (see Hut et al. 1992; Drukier 1996).
 Brandt and Podsiadlowski
(1995) find that $\sim 17 \%$ of binaries remain bound if the kick velocity is
200
km~s$^{-1}$, which is the right order of magnitude to explain the required mass
in dark massive
remnants. However, this is likely to be an upper limit on the retention
fraction because
even systems which remain bound can receive significant centre-of-mass
velocities. Thus,
Drukier (1996),
using the results of Lyne \& Lorimer and Brandt \& Podsiadlowski, has shown
that most globular
clusters would retain $\sim 1-5 \%$ of their neutron stars.
In the
light of this, we should point out that the lack of a low velocity tail in our
distribution is
not an artefact of the $V_{\rm max}$ weighting. Of the 51 pulsars in our sample
with ages less
than $10^7$ years, the lowest transverse velocity is 70 km~s$^{-1}$. If there
are young
pulsars with very small velocities, then they have not been measured yet.
Another possible
complication is the creation of fast (P $<$ 0.1 s) pulsars with initial timing
ages
$> 10^7$ years. The birthrate of such pulsars is not constrained by our
analysis because
of both our age cutoff and the restriction of our analysis to the
early Molonglo and Green Bank surveys.

In conclusion, we have shown that the distribution of young pulsar proper
motions, corrected for
selection effects, is consistent with a characteristic kick velocity at birth
of $\sim$ 250-300
km~s$^{-1}$. We find little evidence for a significant low velocity tail to the
kick distribution.
Our method is robust in its reproduction of the
mean velocity and low velocity shape of the distribution. However, the shape of
the distribution
at velocities $>$ 300 km~s$^{-1}$ is not well constrained by this method
because of poor
statistics.
Our results are in good agreement with the properties of binaries containing
neutron stars
and  pleasantly close to the value inferred from numerical supernova
simulations
(e.g. Burrows, Hayes \& Fryxell 1996, who inferred kicks of $\sim 300 \, \rm km
\, s^{-1}$
from core recoil during the collapse, but neglected asymmetries in the initial
mass distribution,
see Burrows \& Hayes 1996, which could lead to even higher velocities).
Finally, we have shown that the pulsars with long characteristic ages show the
asymmetric
drift, corresponding to a dynamically old population. The velocity distribution
of these
pulsars is affected by their genesis in binaries as well as their subsequent
motion
through the galaxy.

This work was supported by NSF grant AST93-15455 and NASA grant
NAG5-2756. We would like to thank the referee Philip Podsiadlowski
for comments on the original manuscript.

\clearpage

\begin{figure}
\caption{{\bf Sample Definition:}
The filled circles indicate the pulsars that we include in our analysis. The
open circles denote those that are excluded. Since the purpose here is simply
to demonstrate
which pulsars are in the sample discussed, we omit any error bars. Of the three
excluded
pulsars with $ P > 0.1 s$, two are recycled binaries, B0655+64 and B0820+02.}
\label{select}
\end{figure}

\begin{figure}
\caption{{\bf Luminosity and Velocity for the Proper Motion Pulsars:}
We again include all the pulsars with proper motions on this plot. The filled
circles will be the ones to which our analysis applies. The dotted line
indicates a
proper motion of 5 mas~yr$^{-1}$ and flux 4 mJy. This line does not represent a
cutoff over most
of this diagram because making a pulsar brighter at a given distance will move
it to the
right and making a pulsar faster at a given distance will move it up, thus one
can
populate both sides of the line with observable pulsars. Nevertheless, at the
high
luminosity/high velocity end, it should represent the limiting case. It is also
noticeable
that, on average, higher velocity pulsars have higher luminosities, and so will
be overrepresented
in an unweighted sample.}
\label{Lv}
\end{figure}

\begin{figure}
\caption{{ \bf Selection Effects:}
 The solid circles represent those pulsars
detected in the Green Bank surveys. The open circles are the ones
detected in the Molonglo surveys. The dotted lines indicate the approximate
limiting flux as a function of P for $DM$=50 $\rm cm^{-3} \, pc$ and the most
sensitive Molonglo and Green Bank surveys. The dashed lines are for
$DM$=200 $\rm cm^{-3} \, pc$.}
\label{Slim}
\end{figure}

\begin{figure}
\caption{{\bf Detectable Volume for Each Pulsar:}
 Pulsars with small luminosities will only be observable near the galactic
plane, so that
their $V_{\rm max}$ will be spherical (neglecting other selection effects for
the
moment). Pulsars with large luminosities will be observable much further away,
out to the limits
of the disk that such a population, born in the galactic plane, would fill. In
this case, the
spherical $ V_{\rm max}$ will be cut off above the limits of the disk height
given by $ z = V_z
\times
t$.}
\label{Volume}
\end{figure}

\begin{figure}
\caption{{\bf The Weighted Proper Motion Distribution:}
The size of each circle is proportional to the logarithm of the weight accorded
that pulsar. The vertical dotted line indicates the dividing line between what
we
consider `young' and `old' pulsars. The horizontal dashed lines indicate the
range
of values we infer for the mean proper motion of the pulsars in each sample.
The
short horizontal solid line at left indicates the mean proper motion value
quoted by
Lyne \& Lorimer (1994).}
\label{tvw}
\end{figure}

\begin{figure}
\caption{{\bf The Inferred Proper Motion Distribution:}
The cumulative probability distribution $P(V\leq V_{\rm t}$) is shown for
different degrees of correction. The dotted line shows the entire sample with
no
$V_{\rm max}$ weighting and no treatment of upper bounds. The short dashed line
is for only the 66 pulsars with detected proper motions corrected using our
$V_{\rm max}$
correction (i.e., all those with only upper limits were left out).
 The long dashed line is the properly corrected sample (with $V_{\rm max}$
corrections
and Kaplan-Maier estimator) but only those with characteristic ages greater
than $10^7$ years.
 The solid line is the complementary sample of only those pulsars with
characteristic
ages less than $10^7$ years. The 95\% confidence levels for the corrected
distributions
lead to an uncertainty of $\sim 0.05$ in P for velocities $<$ 300 km \,
s$^{-1}$. Above that,
the statistics become uncertain and we cannot say much about the distribution.}
\label{pdis}
\end{figure}

\begin{figure}
\caption{{\bf The Asymmetric Drift:}
The solid squares indicate a positive $V_{\rm y}$ and open squares indicate a
negative $V_{\rm y}$. The
circled points indicate binaries. The horizontal dashed line indicates the
order of magnitude of
the sun's motion within its local standard of rest, and velocities below this
will experience
a contamination of
the asymmetric drift. The vertical dashed line is at 4 $\times 10^6$ years, and
represents an approximate division between ``young'' and ``old'' pulsars,
namely those that show the asymmetric
drift and those that don't. Of the young pulsars, only 33\% have positive
$V_{\rm y}$, while 91\% of the
old pulsars have positive $V_{\rm y}$. The dashed line indicates the evolution
of the mean asymmetric
drift velocity obtained from the calculation in
section~\protect{\ref{Evolve}}.}
\label{asd}
\end{figure}

\begin{figure}
\caption{{\bf Determining the Birth Velocity Distribution:}
The heavy solid line is the corrected proper motion distribution, with the
formal uncertainty
shown by the shaded region. The short dashed
line is the proper motion distribution obtained for a distribution given by two
delta functions
at 250 km/s and 1000 km/s, in random directions. This distribution and the
maxwellian given by
(\protect{\ref{pkick}}) (thin solid line) are consistent with the data.
 The alternating and
long dashed lines represent the results of Paczynski (1990) and Lyne \& Lorimer
(1994), neither
of which are fully consistent with the data.
}
\label{vdis}
\end{figure}

\begin{figure}
\caption{{\bf Monte Carlo Pulsars:} The right half of this projection onto
the galactic plane is for the initial positions of the pulsars. The mirror
images
of the positions of the same pulsars after $10^9$ years is shown on the left
hand side. The circles indicate the volumes sampled by the `observations'.}
\label{Observe}
\end{figure}

\begin{figure}
\caption{{\bf Velocity Distribution Evolution:} The solid lines, labelled by
log(age), are the
time-dependant transverse
velocity distributions produced by the initial three dimensional
distribution labelled $ V_{\rm 0}$ (dotted line).
 The dashed line indicates the observed `young' ($<10^7$ years) pulsar proper
motion distribution.
 The heavy line indicates the observed `old' ($>10^8$ years) proper motion
distribution.
The line is solid where the contribution is from a binary and dotted when the
corresponding pulsar is isolated.}
\label{vevol}
\end{figure}

\begin{figure}
\caption{The four panels (all to the same scale) show the radial ($V_x$) and
transverse ($V_y$)
velocities in the observer frame at four different ages. After $10^6$ years the
velocity distribution
is isotropic because only locally born pulsars are seen. After $10^7$ years an
excess of radially
outward motion is seen due to the net flux of pulsars moving out from the inner
parts of the
galaxy. After $10^8$ years only pulsars from the lower velocity tail are seen
and these clearly
exhibit the asymmetric drift.}
\label{simasd}
\end{figure}

\label{lastpage}
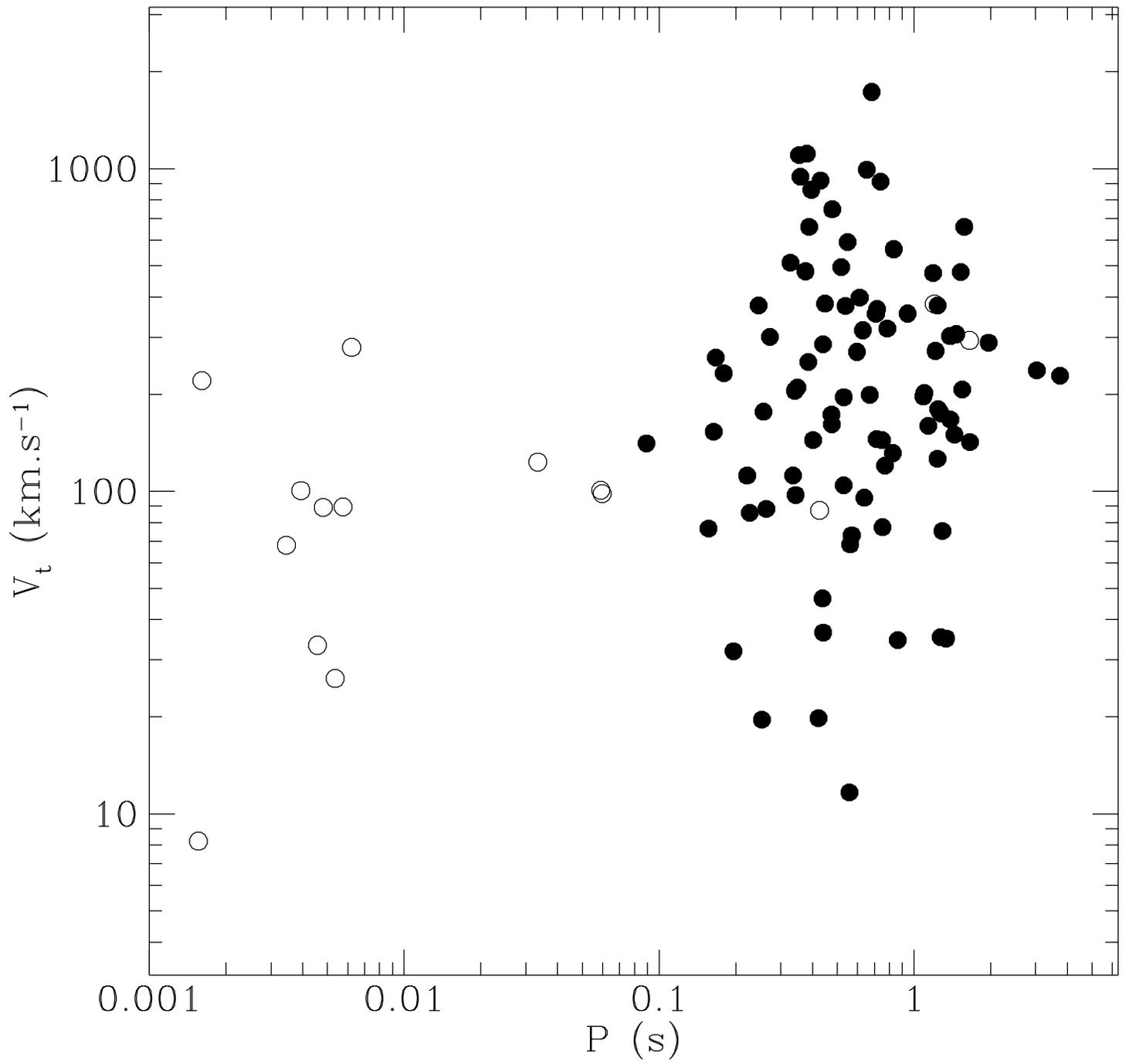
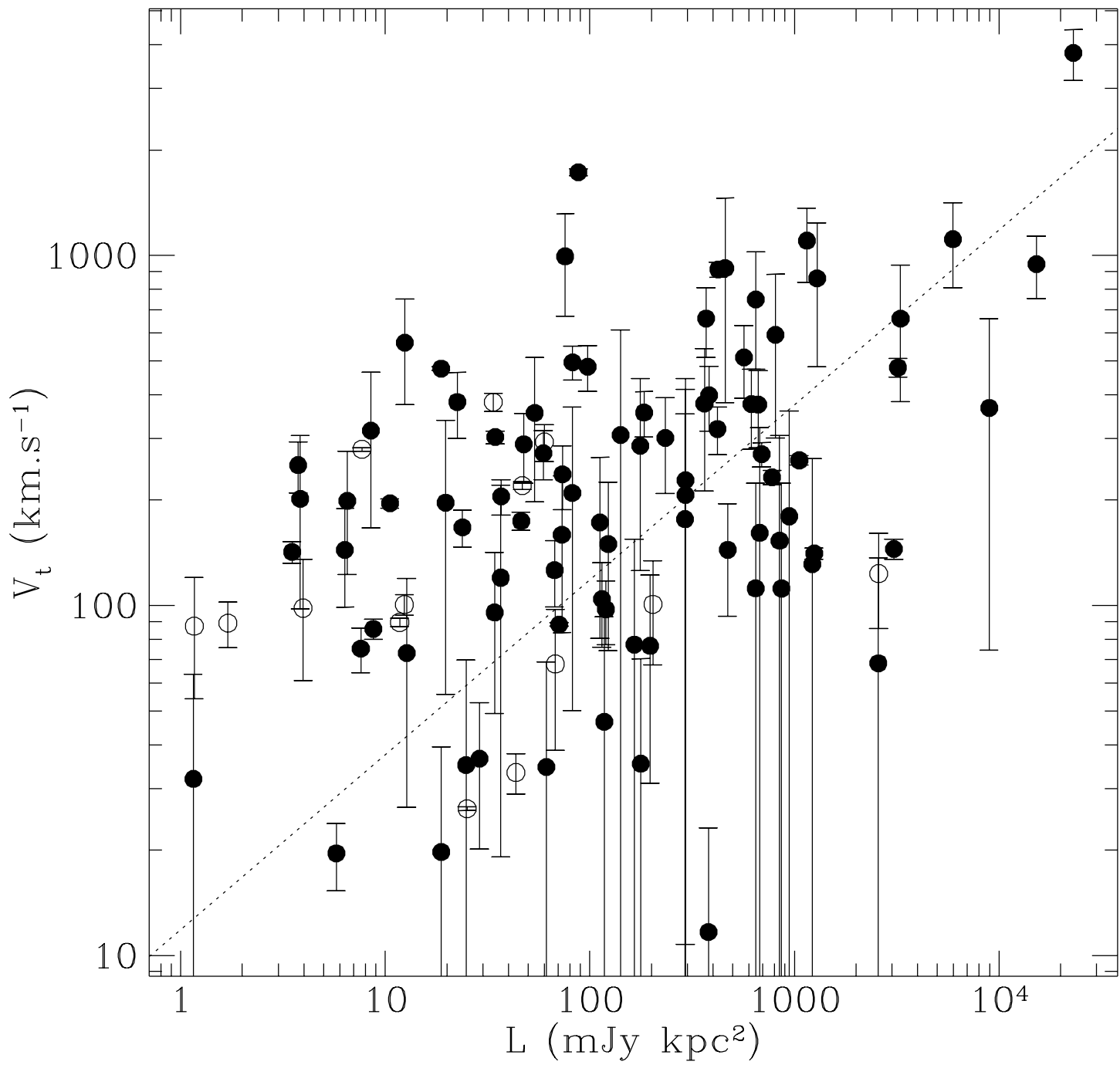
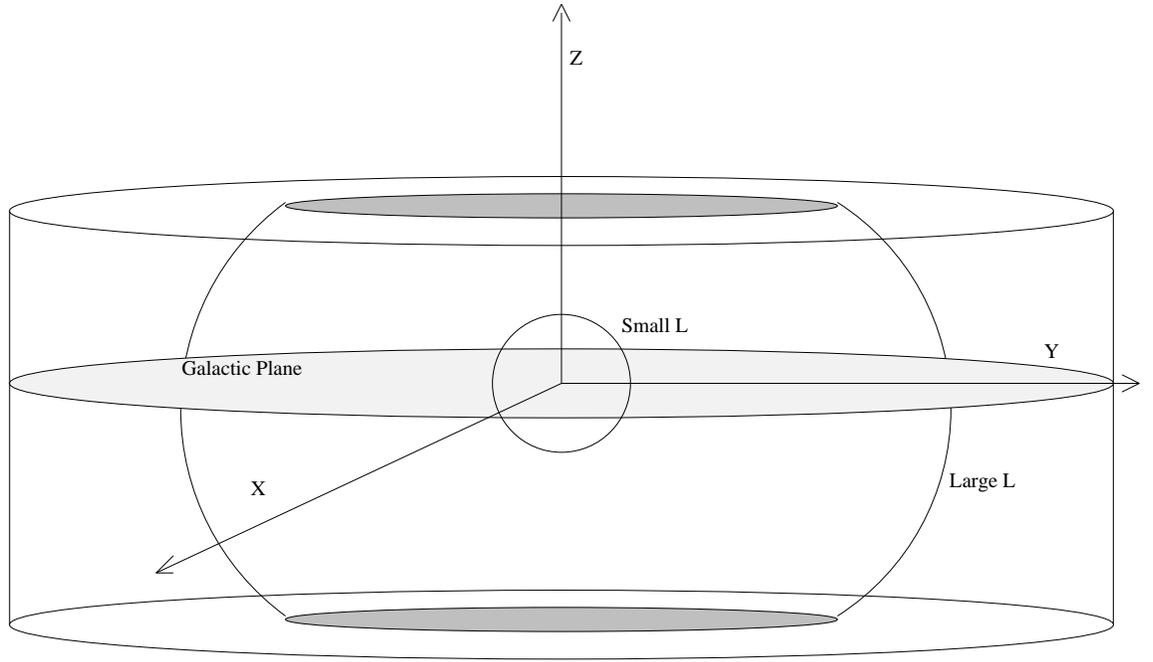
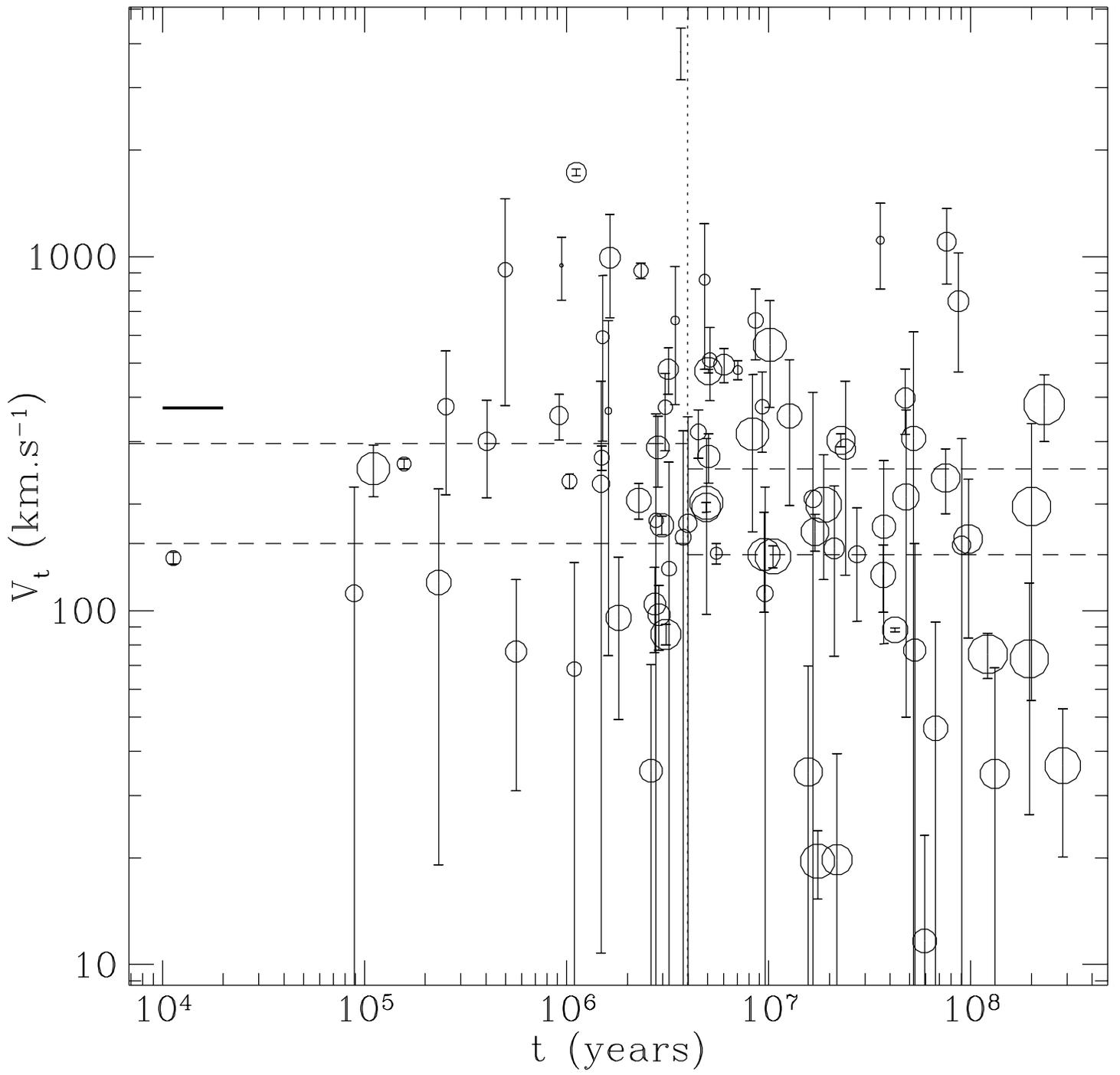
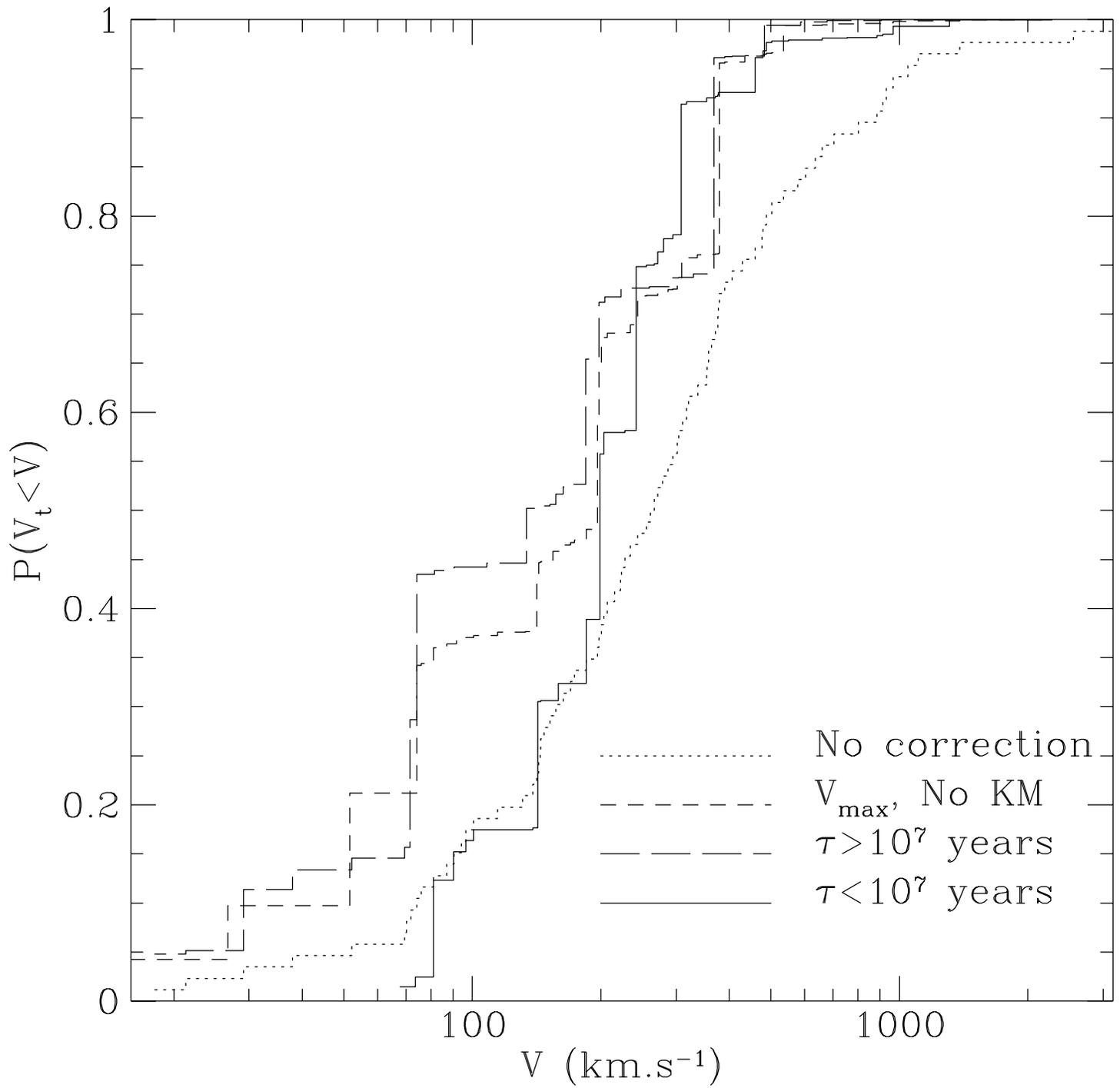
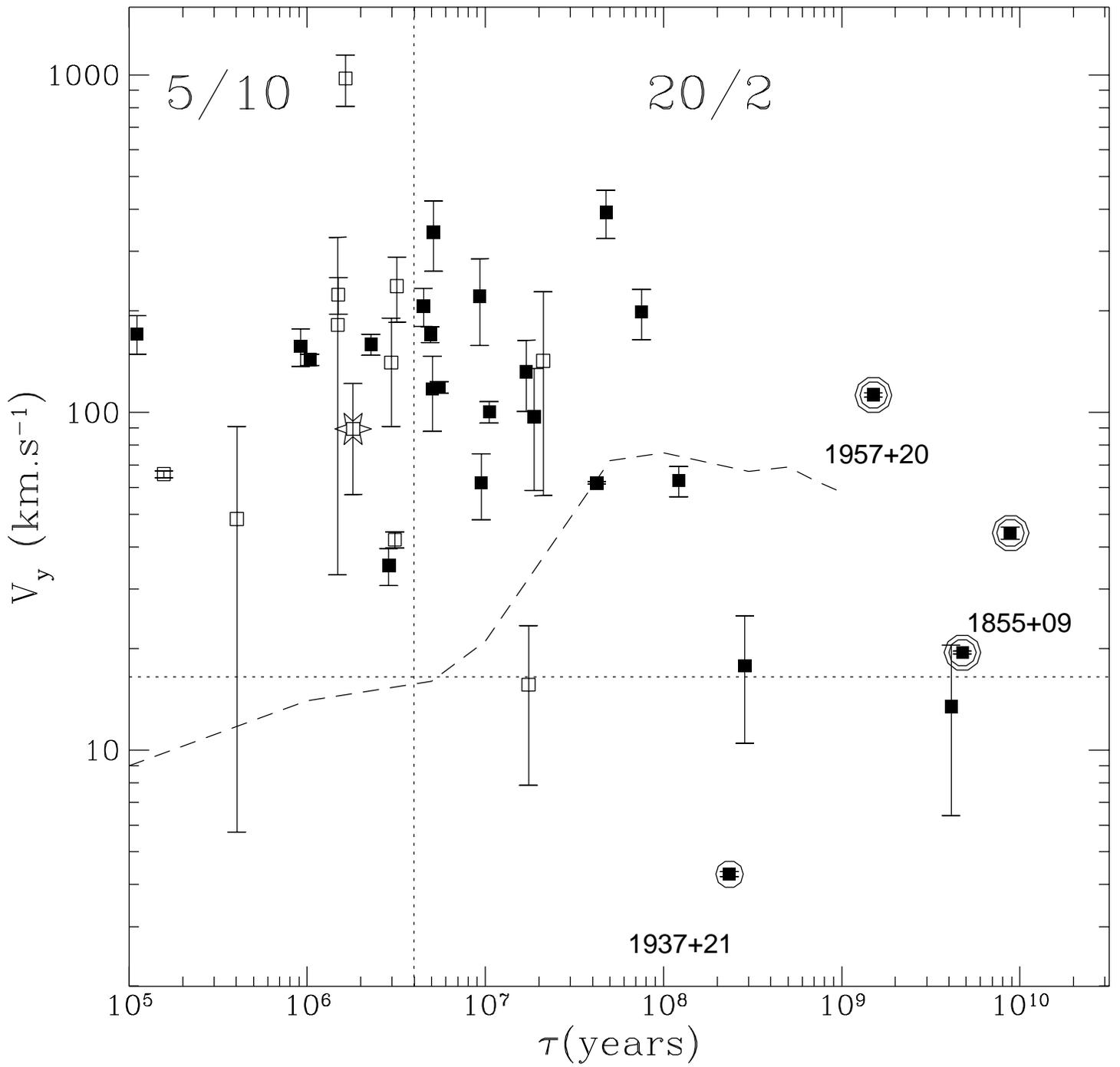
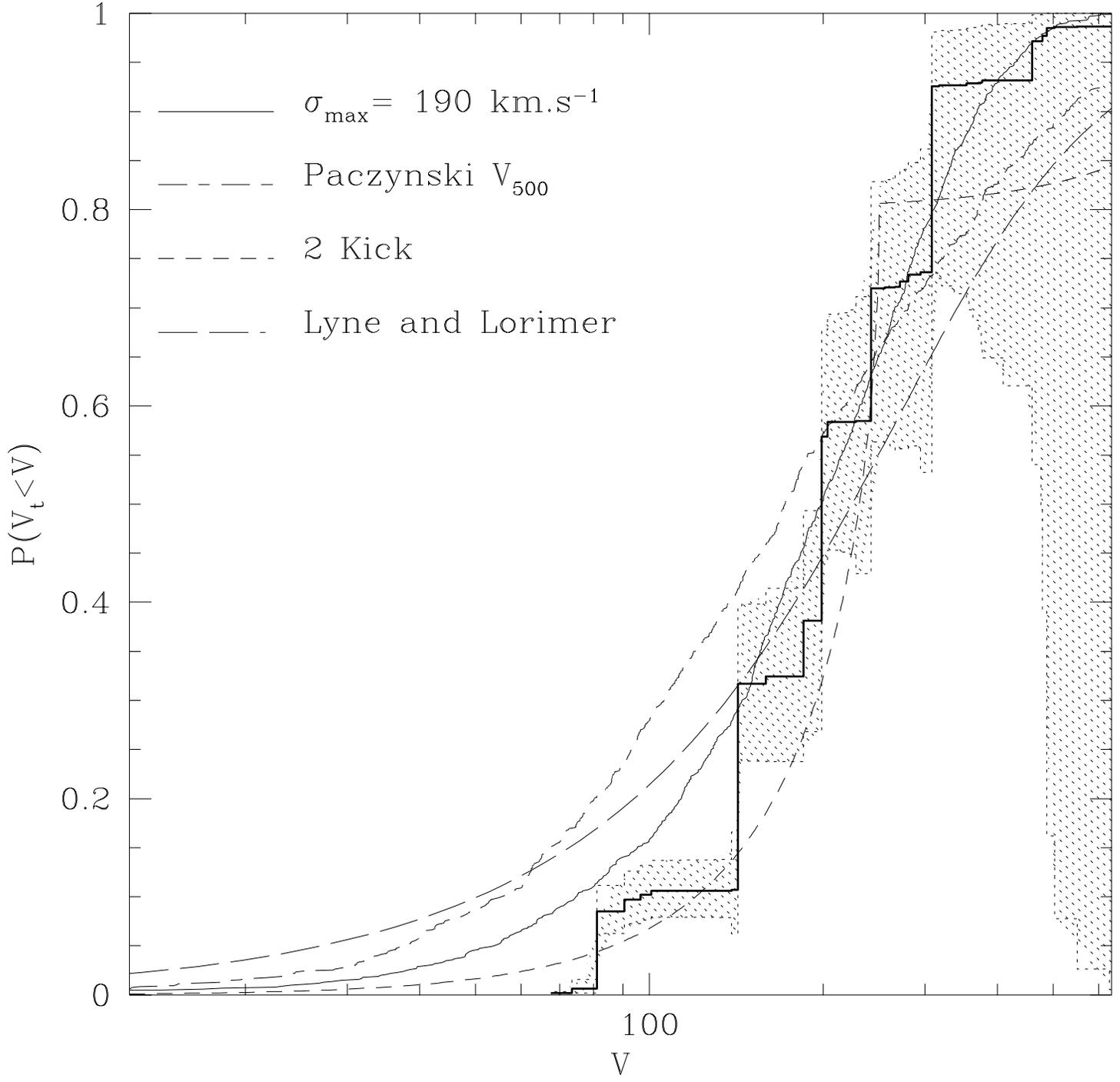
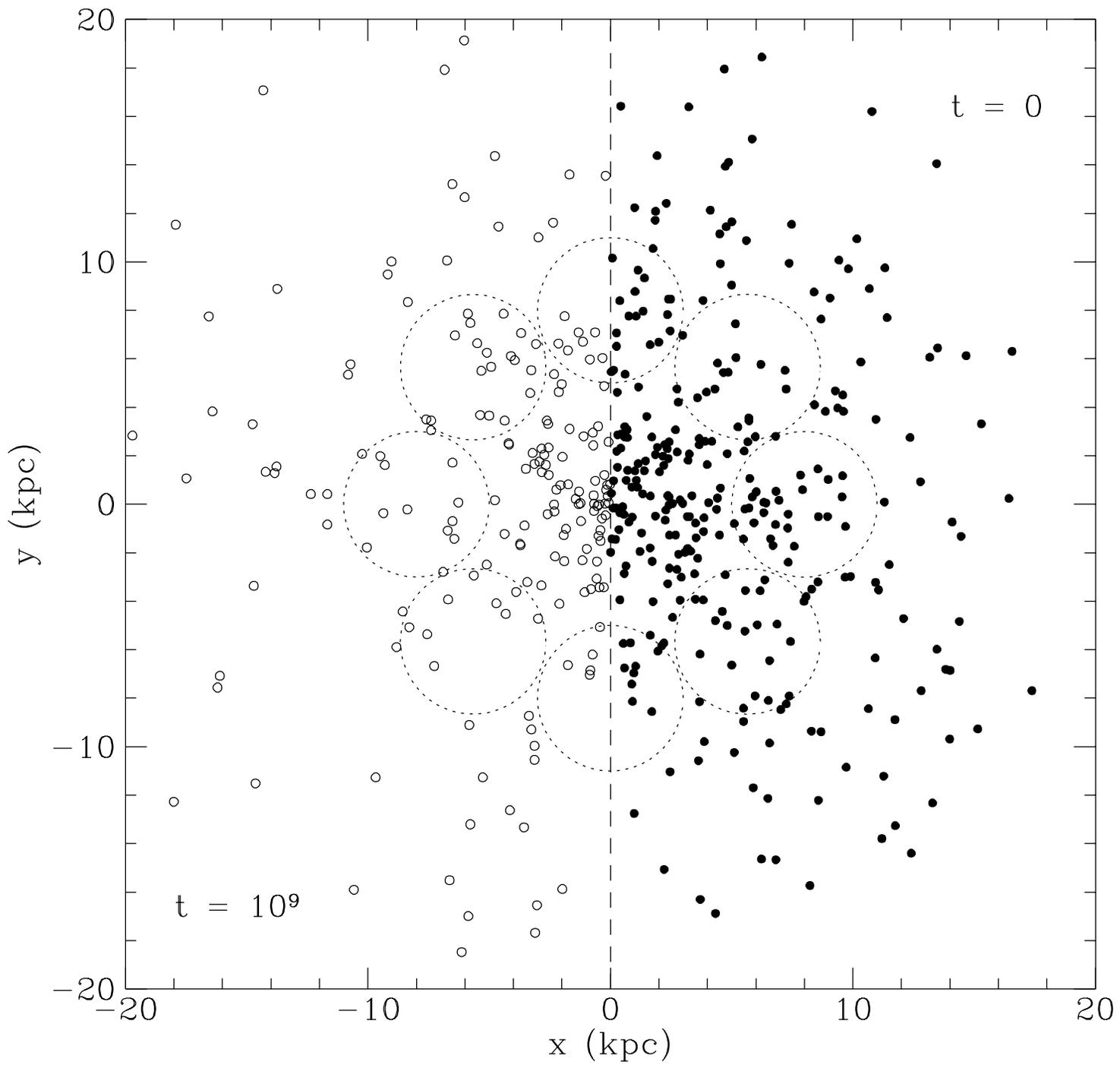
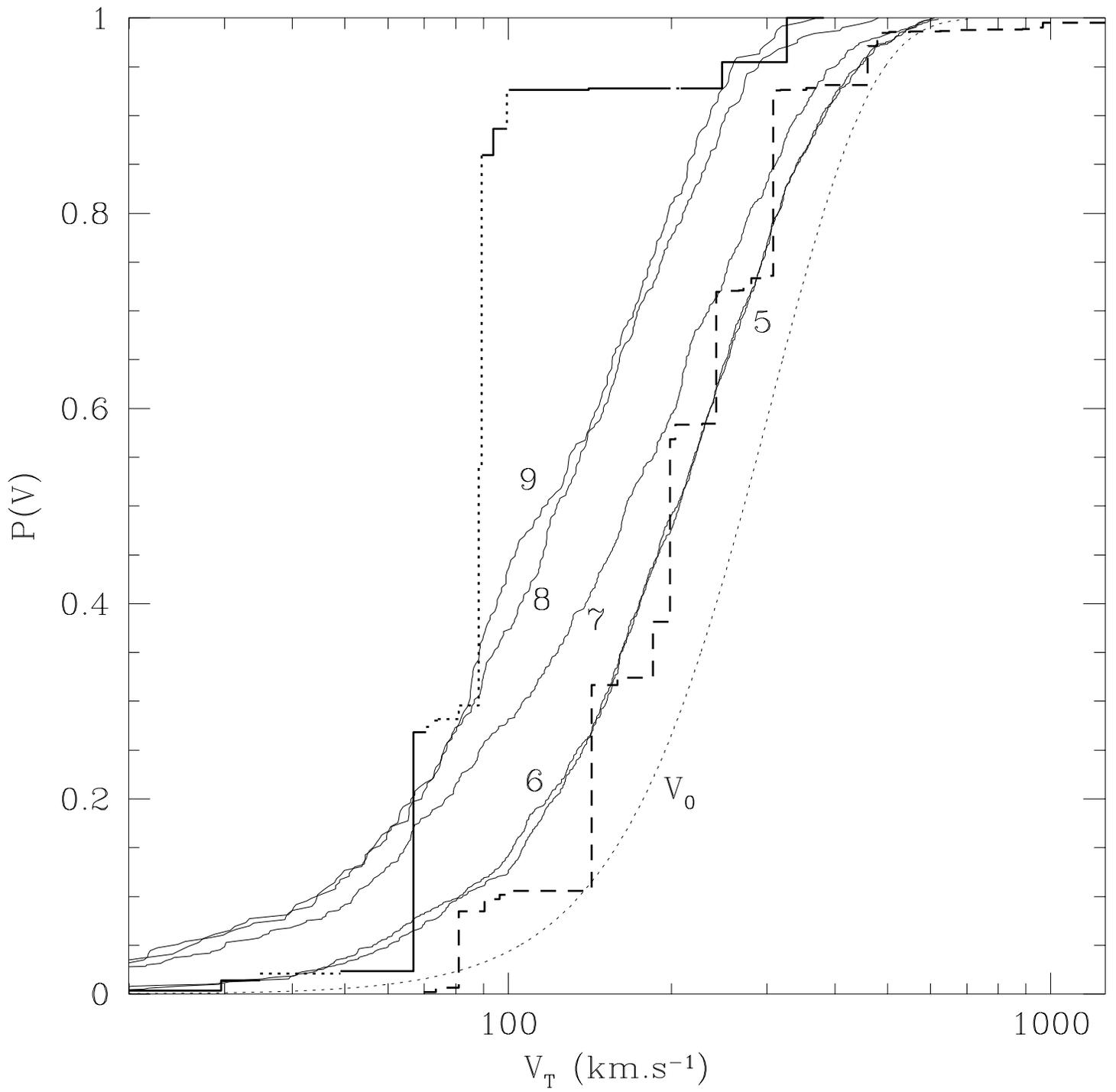
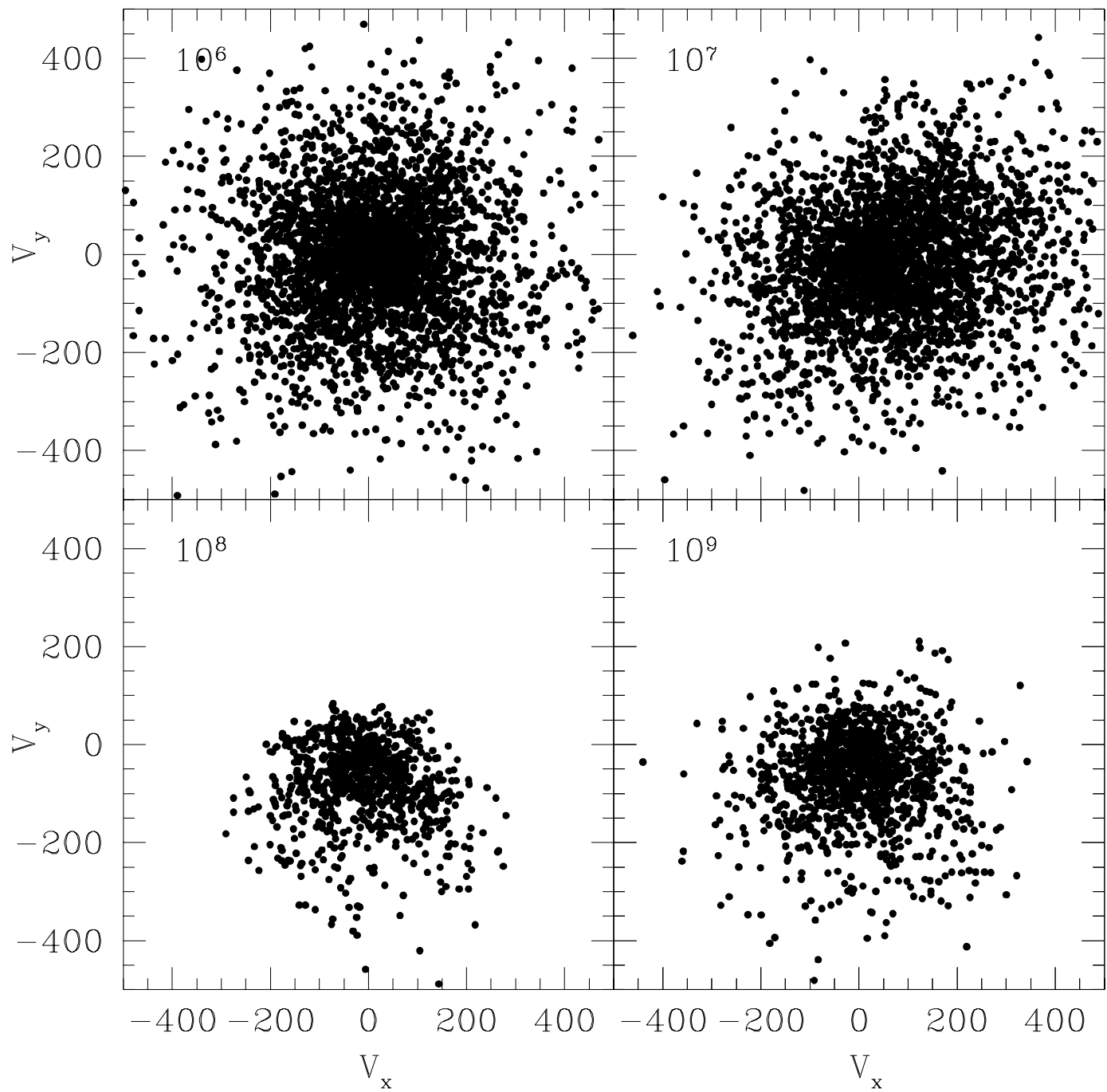
\end{document}